\input{aipcheck}

\documentclass[
    ,final     
  ]
  {aipproc}

\layoutstyle{8x11single}
\newcommand{\nc}{\newcommand}%
\nc{\beq}{\begin{eqnarray}}
\nc{\eeq}{\end{eqnarray}}
\nc{\ba}{\begin{array}}
\nc{\ea}{\end{array}}
\nc{\la}{\label}
\nc{\no}{\nonumber}
\nc{\btab}{\begin{tabular}}
\nc{\etab}{\end{tabular}}
\nc{\ci}{\cite}
\nc{\se}{\section}
\nc{\ti}{\tilde}
\nc{\rs}{\slash\hspace{-0.2cm}\partial}

\begin{document}

\title{On the contribution of plasminos to the shear viscosity of a hot and dense Yukawa-Fermi gas\footnote{Poster presented by F. Taghinavaz at ``XIth Quark Confinement and the Hadron Spectrum'', Saint-Petersburg, Russia, 8-12 September 2014.}}


\author{N. Sadooghi\footnote{Corresponding author: sadooghi@physics.sharif.ir}~}{address={
Department of Physics, Sharif University of Technology, P. O. Box 11155-9161, Tehran, Iran
}}

\author{F. Taghinavaz\footnote{taghinavaz@physics.sharif.ir}}{address={
Department of Physics, Sharif University of Technology, P. O. Box 11155-9161, Tehran, Iran
}}

\begin{abstract}
Using the standard Green-Kubo formalism, we determine the shear viscosity $\eta$ of a hot and dense Yukawa-Fermi gas. In particular, we study the effect of particle and plasmino excitations on thermal properties of the fermionic part of the shear viscosity, and explore the effects of thermal corrections to particle masses on bosonic and fermionic shear viscosities, $\eta_{b}$ and $\eta_{f}$. It turns out that the effects of plasminos on $\eta_{f}$ become negligible with increasing (decreasing) temperature (chemical potential).
\end{abstract}

\maketitle


\section{Introduction}
Exploring the nature of the medium created after relativistic heavy-ion collisions has attracted much attention in recent years~\ci{AR.2005}. It has been shown, that after thermalization~\ci{Heinz:2013th} this medium undertakes a collective evolution. Theoretically, this indicates the dominance of the long wave-length limit of the considered theory. The corresponding thermal medium can therefore be described by relativistic hydrodynamics or, in general, relativistic viscous hydrodynamics. The latter is characterized by a number of transport coefficients. In the present work, we will focus on the shear viscosity, which measures the resistance of a fluid against a lateral flow~\ci{Landau.F}. In general, there are two different approaches for determining transport coefficients: The Kinetic Theory (KT) and the Green-Kubo formalism. The former is based on solving the Boltzmann equation in the linear regime in which all disturbances are small enough compared to the average values~\ci{Lifshitz.10}, and the latter is based on the theory of linear response. In the framework of KT, certain divergences appear in the perturbative computation of transport coefficients. In the massless theories and at high temperature, these divergences arise from small-angle scatterings~\ci{heiselberg}, and can be remedied by including Hard Thermal Loop (HTL) corrections to the internal propagators of the fermionic modes~\ci{AMY2}.  In the Green-Kubo approach, in contrast, these divergences appear as a resummation of an infinite number of diagrams, which has one and the same order of coupling constant~\ci{Jeon94}. In this way, the conventional perturbative treatment breaks down in the long wave-length (hydrodynamics) limit of the considered theory~\ci{Arnold97}.

In the present work, we have used the Green-Kubo formalism to determine the dependence of the shear viscosity of a hot and dense Yukawa-Fermi gas on temperature (T), chemical potential ($\mu$) as well as on bosonic and fermionic masses, $m_{b}$ and $m_{f}$.\footnote{See \cite{Taghinavaz2014} for a more detailed analysis.}
Since we are interested on the regimes of low and moderate temperature, where particle masses cannot be ignored, we shall not be worried about the breakdown
of our perturbative treatment. As aforementioned, the latter is caused by the appearance of certain divergences in the chiral limit. Considering the bosonic and fermionic masses
enables us to include HTL corrections via an appropriate modification of particle masses in the thermal medium. The novelty in the present work is considering the contribution
of particle and plasmino modes ~\ci{Weldon82} in the thermal medium, and studying, in particular, their contributions to the fermionic part of the shear viscosity. We will
show that these contributions significantly affect the shear viscosity, in particular at low and moderate temperatures \cite{Taghinavaz2014}.
\section{The Method}
\begin{figure}
  \includegraphics[height=.08\textheight]{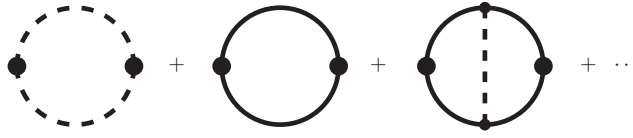}
  \caption{Dashed and solid lines denote dressed  bosonic and fermionic two-point Green functions, respectively. According to the arguments presented in the main text, we limit ourselves only to the first two diagrams \cite{Taghinavaz2014}.}\label{skeleton}
\end{figure}

Our approach is similar to what is performed in~\ci{Lang2012}. To start,
let us introduce the Green-Kubo type formula for the shear viscosity
\beq\la{21}
\eta= \frac{\beta_{s}}{10} \int d^3x~\int_{-\infty}^{t} dt' ( \ti{\pi}^{\mu \nu}(0), \ti{\pi}^{\mu \nu}(x',t')).
\eeq
Here,  $\beta_s\equiv\gamma \beta$ is the inverse proper temperature, with  $\gamma=\frac{1}{\sqrt{1-v^{2}}}$ and $\beta=\frac{1}{T}$, and $\ti{\pi}^{\mu \nu}$ is the traceless part of energy-momentum tensor, $T^{\mu\nu}$,
\beq\label{22n}
\ti{\pi}^{\mu \nu}= (\Delta^{\nu \sigma}\Delta^{\mu \rho} + \Delta^{\nu \rho}\Delta^{\mu \sigma}- \frac{2}{3}\Delta^{\mu \nu}\Delta^{\rho \sigma})T_{\rho \sigma}.
\eeq
To determine $\eta$ from (\ref{21}) in a diagrammatic manner, we adopt an appropriate skeleton expansion, which is demonstrated in Fig. \ref{skeleton}. 
Here, the lines correspond to dressed two-point Green functions (TPGF) of bosons (dashed lines) and fermions (solid lines).
In (\ref{22n}),  $T_{\rho\sigma}$ is the energy-momentum tensor of the underlying quantum field theory, and $\Delta^{\mu \nu}$ is an operator which project any quantity in hypersurface orthogonal to the direction of fluid velocity $u^{\mu}$. It is defined by $\Delta^{\mu\nu}\equiv g^{\mu\nu}-u^{\mu}u^{\nu}$, where the metric $g^{\mu \nu}=\mbox{diag}(+,-,-,-)$.
\par
As it turns out \cite{Lang2012, Taghinavaz2014}, $\eta$ is sensitive to low frequency limit of the considered theory. It is given by
\beq\la{23}
\eta = \frac{i}{10} \frac{d}{dp_0} \Pi_{R}(p_0)\bigg|_{p_0 \to 0},
\eeq
where $\Pi_{R}(p_0)$ is the Fourier transform of the retarded correlator of two traceless operators $\tilde{\pi}^{\mu\nu}$ from (\ref{22n}). According to the above formulation, the shear viscosity $\eta$ can be determined by adopting a specific theory, deriving the traceless part of its energy-momentum tensor from (\ref{22n}) and inserting it into (\ref{23}). In the present work, we will select the Yukawa theory and will determine the corresponding bosonic and fermionic shear vicosities. The Yukawa theory has the well-known Lagrangian
\beq
\mathcal{L}= \bar{\psi}(i ~\rs - m_{f}) \psi + \frac{1}{2} \partial_{\mu}\phi \partial^{\mu}\phi - \frac{1}{2} m_{b}^{2} \phi^2 + g \bar{\psi}\psi \phi.
\eeq
The corresponding energy-momentum tensor of this Lagrangian reads
\beq
T_{\mu \nu}= i \bar{\psi} \gamma_{\mu}\partial_{\nu} \psi + \partial_{\mu}\phi \partial_{\nu}\phi - \mathcal{L} g_{\mu \nu}.
\eeq
Any retarded TPGF of this theory can be related to the corresponding spectral function, $\rho$, via the  K$\ddot{a}$llen-Lehmann representation
\beq\label{26}
G_{T}(\textbf{p}, \omega_{n}) = \int_{-\infty}^{+\infty} \frac{d\omega}{2\pi} \frac{\rho(\textbf{p}, \omega)}{\omega + i\omega_n}.
\eeq
Using (\ref{26}), the spectral function can be shown to be proportional to the imaginary part of the corresponding retarded TPGF. Similarly, the real part of each retarded TPGF can be derived using Kramers-Kronig relation. In this way, the shear viscosity is expressed in terms of the real and imaginary parts of retarded bosonic and fermionic TPGFs.
\par
At this stage some remarks are in order: For the bosonic particles, because of certain symmetry properties \cite{kapusta}, we have in general no ambiguity to determine the spectral function, $\rho_{b}$. It is given by \cite{Lang2012, Taghinavaz2014}
\begin{eqnarray}\label{27n}
\rho_{b}(\mathbf{p},\omega)=\frac{4\omega\Gamma_{b}(\mathbf{p},\omega_{b})}{[\omega^{2}-E_{b}^{2}(\mathbf{p},\omega_{b})-\Gamma_{b}^{2}(\mathbf{p},\omega_{b})]^{2}+4\omega_{2}\Gamma_{b}^{2}(\mathbf{p},\omega_{b})},
\end{eqnarray}
where $\omega_{b}^{2}=\mathbf{p}^{2}+m_{b}^{2}$, $E_{b}^{2}(p)=\omega_{b}^{2}+\mbox{Re}[\Sigma_{R}^{b}(p)]$ and $\Gamma_{b}(p)=-\frac{1}{2p_{0}}\mbox{Im}[\Sigma_{R}^{b}(p)]$. Here, $\Sigma_{R}^{b}$ is the retarded bosonic TPGF. As concerns the fermionic  spectral function $\rho_{f}$, however, certain ambiguity occurs. As it turns out, $\rho_{f}$ includes both particles and antiparticles. In the chiral limit, it is given by~\ci{Gagnon2007}
\beq\la{22}
\rho_{f}=\frac{\Gamma_{k}}{(\omega - E_{k})^{2}+ \Gamma_{k}^{2}/2} h_{+}(\hat{k})+\frac{\Gamma_{k}}{(\omega + E_{k})^{2}+\Gamma_{k}^{2}/2} h_{-}(\hat{k}),
\eeq
where the projector $h_{+}(\hat{k})$ and $h_{-}(\hat{k})$ correspond to particle and antiparticle contribution to $\rho_{f}$, respectively. Moreover, $\Gamma_{k}$ is the decay width of both fermions and antifermions. Let us notice that at high temperature, where $m_{f}\ll T$, there is no difference between the decay widths corresponding to fermions and antifermions. However, at moderate $T$, where particle masses turn out to have nontrivial effects on the quantities appearing in $\rho_{f}$ in (\ref{22}), we have to distinguish between the decay widths of fermions and antifermions. One of these effects is the appearance of new degrees of freedom, the so called plasminos~\ci{Weldon89}. Plasminos are similar to antiparticles, in the sense that both plasminos and antiparticles have opposite chirality and helicity. Their difference arises in their corresponding dispersion relations, once the contributions of T-dependent part of the radiative corrections are taken into account \ci{Weldon89}. The expression (\ref{22}) for $\rho_{f}$ turns out to be therefore invalid for $m_{f}\approx T$. Here, the difference between particles and plasminos implies  a more accurate form of $\rho_{f}$ \cite{Taghinavaz2014}, 
\beq \la{24}
\rho_{f}(\textbf{p},\omega) = \frac{2 \Gamma_{+}(\textbf{p},\omega_{f})}{(\omega - E_{+}(\textbf{p},\omega_{f}))^{2} + \Gamma_{+}^{2}(\textbf{p},\omega_{f})} \hat{g}_{+} - \frac{2 \Gamma_{-}(\textbf{p},\omega_{f})}{(\omega + E_{-}(\textbf{p},\omega_{f}))^{2} + \Gamma_{-}^{2}(\textbf{p},\omega_{f})} \hat{g}_{-},
\eeq
where $\omega_{f}^{2}=\textbf{p}^{2}+ m_{f}^{2}$ and $$\hat{g}_{\pm}(\textbf{p},\omega_{f})= \frac{\gamma_0 \omega_{f}\mp(\gamma.\textbf{p}-m_{f})}{2\omega_f}.$$
In (\ref{24}), $E_{\pm}(\textbf{p},\omega_{f})$ and $\Gamma_{\pm}(\textbf{p},\omega_{f})$ denote the real and imaginary parts of the fermionic TPGF, respectively. They are defined by
\beq
E_{\pm}(\textbf{p},\omega_{f}) &\equiv& \omega_{f} \pm \frac{1}{2} tr(\hat{g}_{\pm}(\textbf{p},\omega_{f}) \mbox{Re}[\Sigma_{R}^{f}(\textbf{p},\omega_{f})]),\nonumber\\
\Gamma_{\pm}(\textbf{p},\omega_{f})&\equiv& \pm \frac{1}{2} tr(\hat{g}_{\pm}(\textbf{p},\omega_{f}) \mbox{Im}[\Sigma_{R}^{f}(\textbf{p},\omega_{f})]).
\eeq
Let us notice that in (\ref{24}), comparing to (\ref{22}), particle masses as well as the difference between the decay widths corresponding to particles and plasminos are taken into account \cite{Taghinavaz2014}. Here, the imaginary part of bosonic and fermionic TPGT is computed using the real-time formalism (RTF) of finite temperature field theory (TFTF)~\ci{Semenof, Adas}. In what follows, we will present the final results for $\Gamma_{b}, \Gamma_{\pm}$ as well as bosonic and fermionic shear viscosities $\eta_{b}$ and $\eta_{f}$.
\begin{figure}[!t]
\includegraphics[height=.25\textheight]{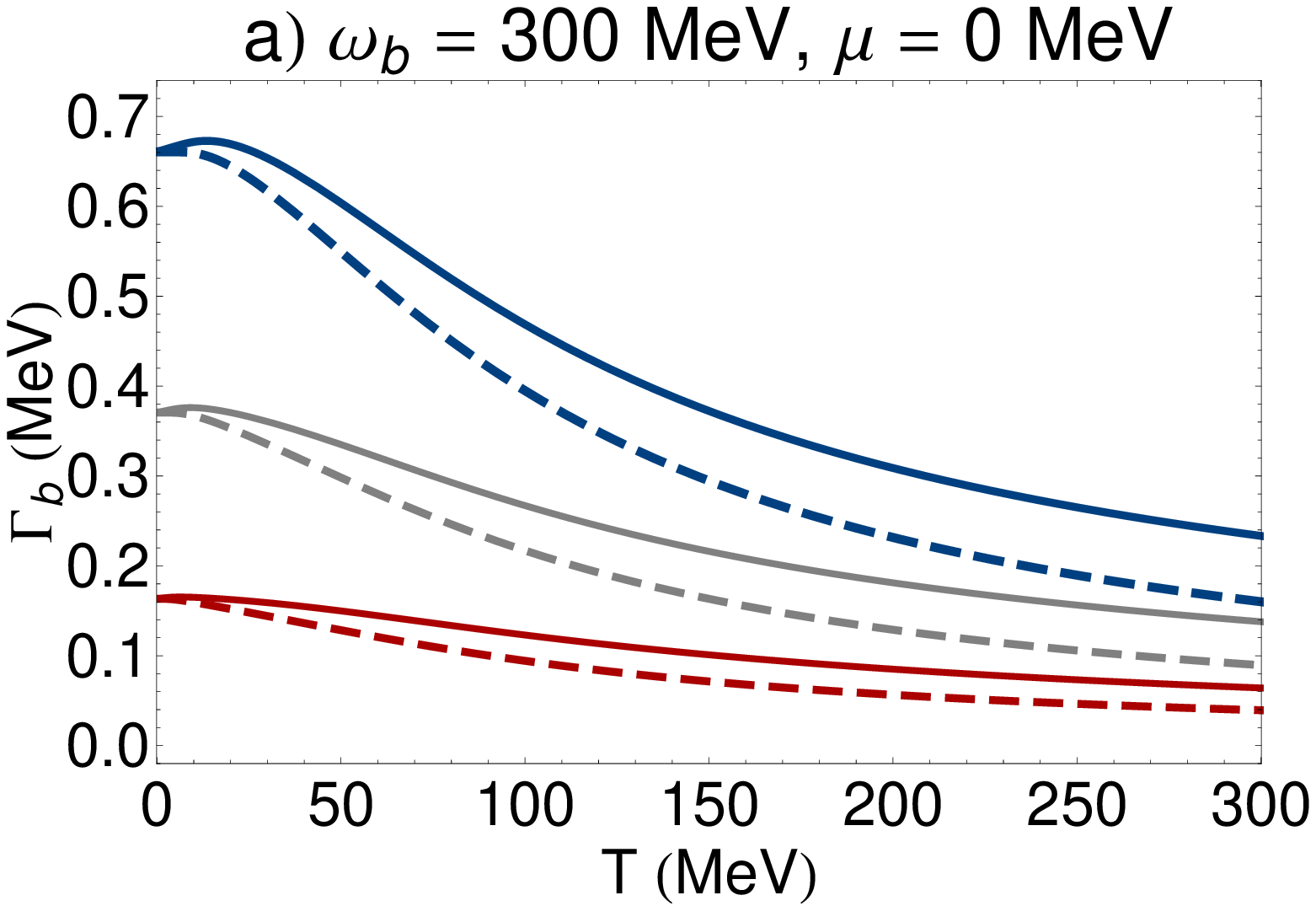}
\includegraphics[height=0.25\textheight]{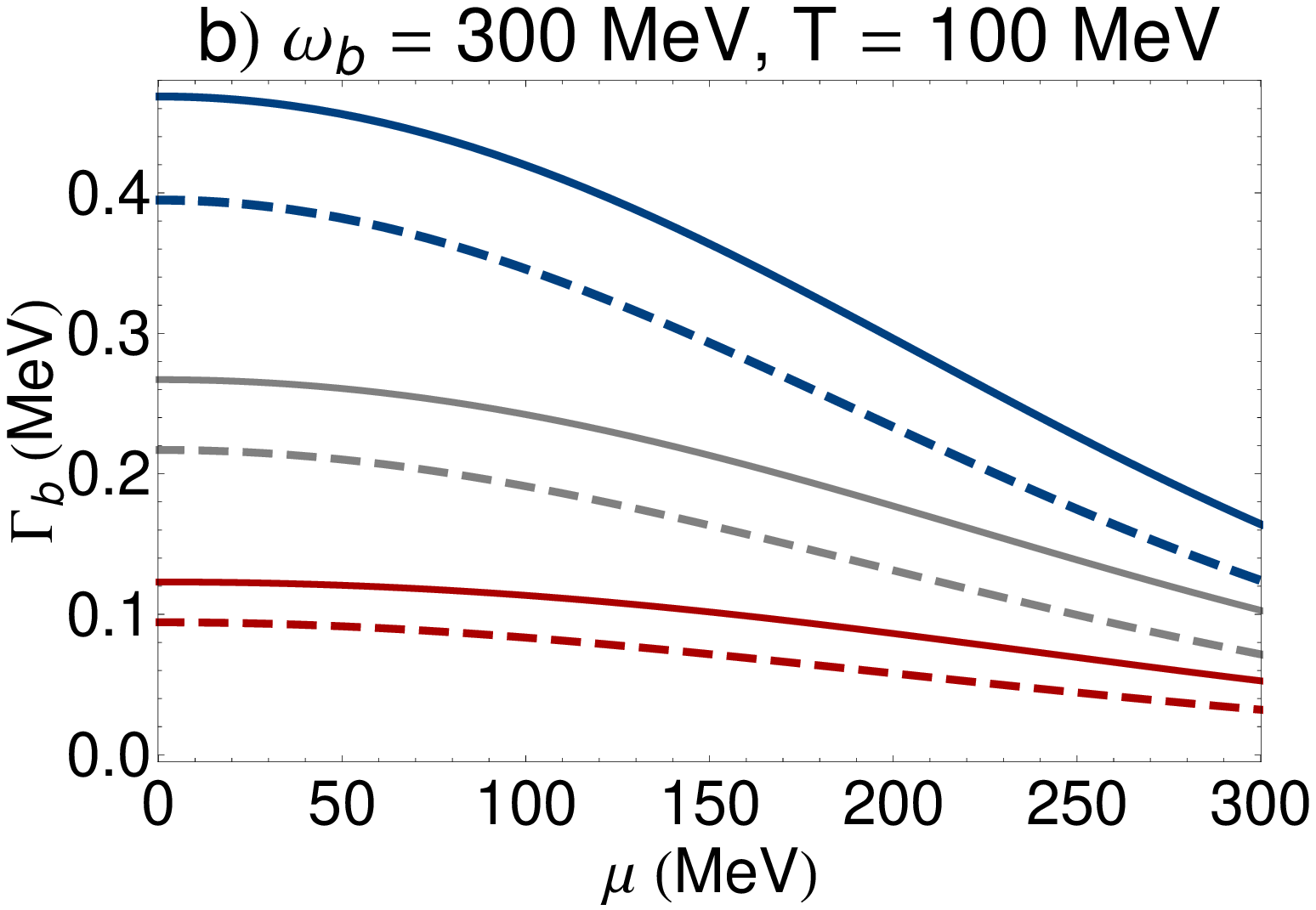}
\caption{(color online). (a) The T  dependence of $\Gamma_{b}$ for $\omega_{b}=300$ MeV and (b) the $\mu$ dependence of $\Gamma_{b}$ for $\omega_{b}=300$ MeV. The dashed lines correspond to $\Gamma_{b}$ including the constant mass contributions of bosons, while the solid lines correspond to the same quantity including thermal corrections of bosonic masses described by (\ref{31}). The red, gray and blue lines correspond to $m_{b}^{0}=100,150,300$ MeV and $m_{f}^{0}=5$ MeV. Here, the Yukawa coupling $g=0.5$ is used~\ci{Taghinavaz2014}.}
\la{gammabt}
\end{figure}
\begin{figure}[!b]
\includegraphics[height=0.25\textheight]{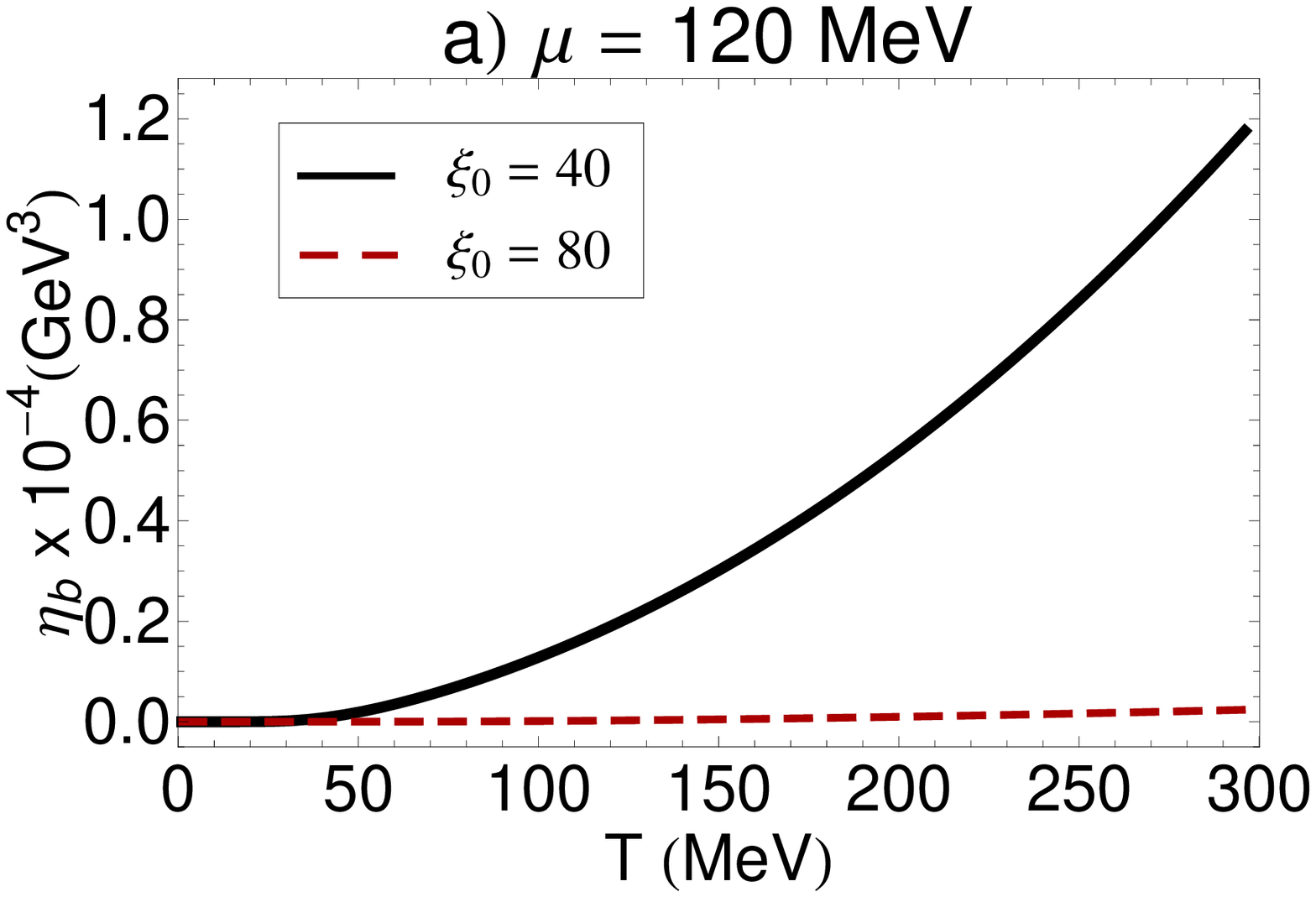}
\includegraphics[height=0.25\textheight]{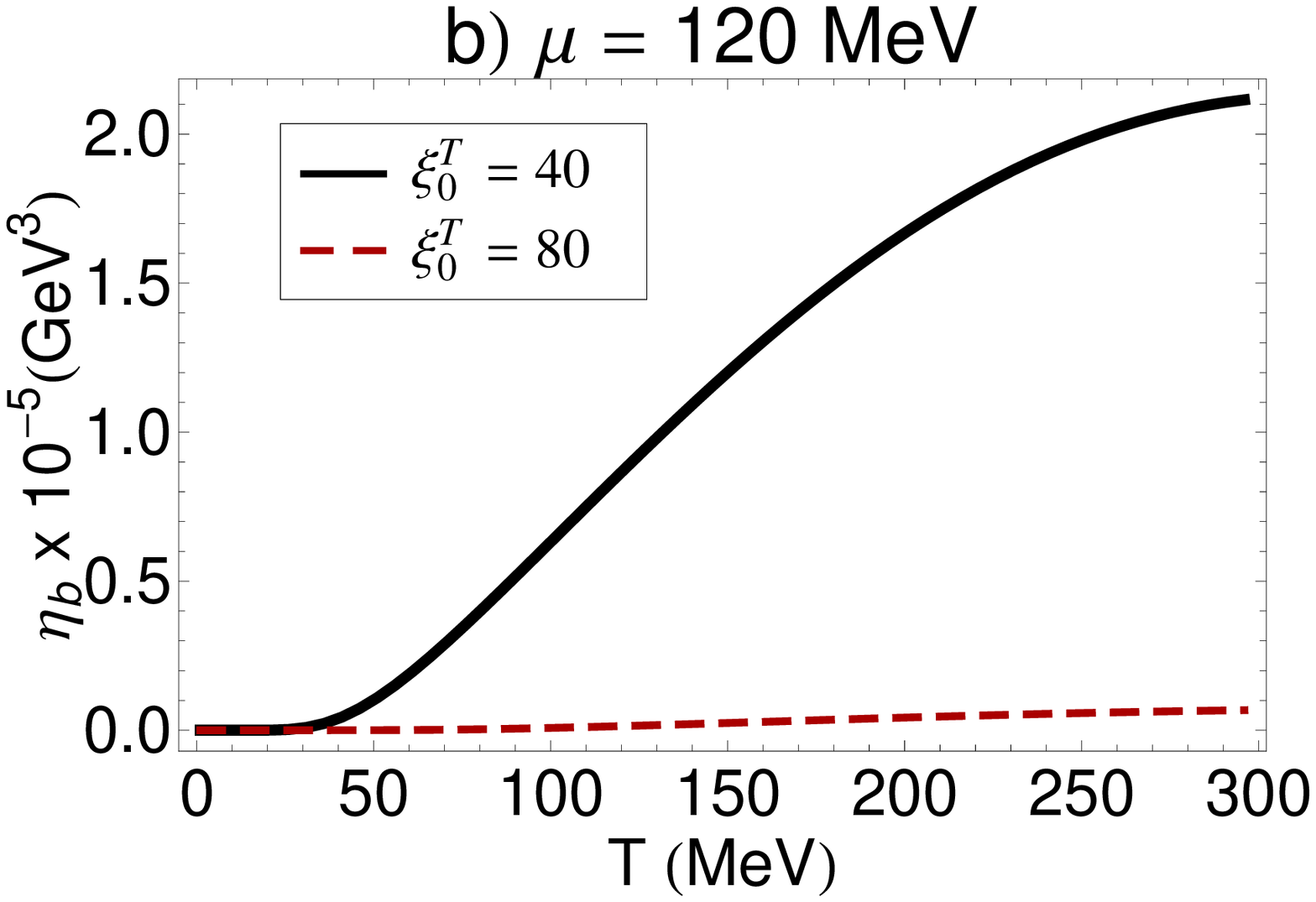}
\caption{(color online).  The T dependence of $\eta_{b}$ is plotted for $\mu=120$ MeV for (a) (T, $\mu$)-independent $\xi_{0}= 40,80$ and (b) (T, $\mu$)-dependent $\xi_{0}^{T}= 40,80$~\ci{Taghinavaz2014}.}
\la{etabt}
\end{figure}
\begin{figure}[!t]
\includegraphics[height=0.25\textheight]{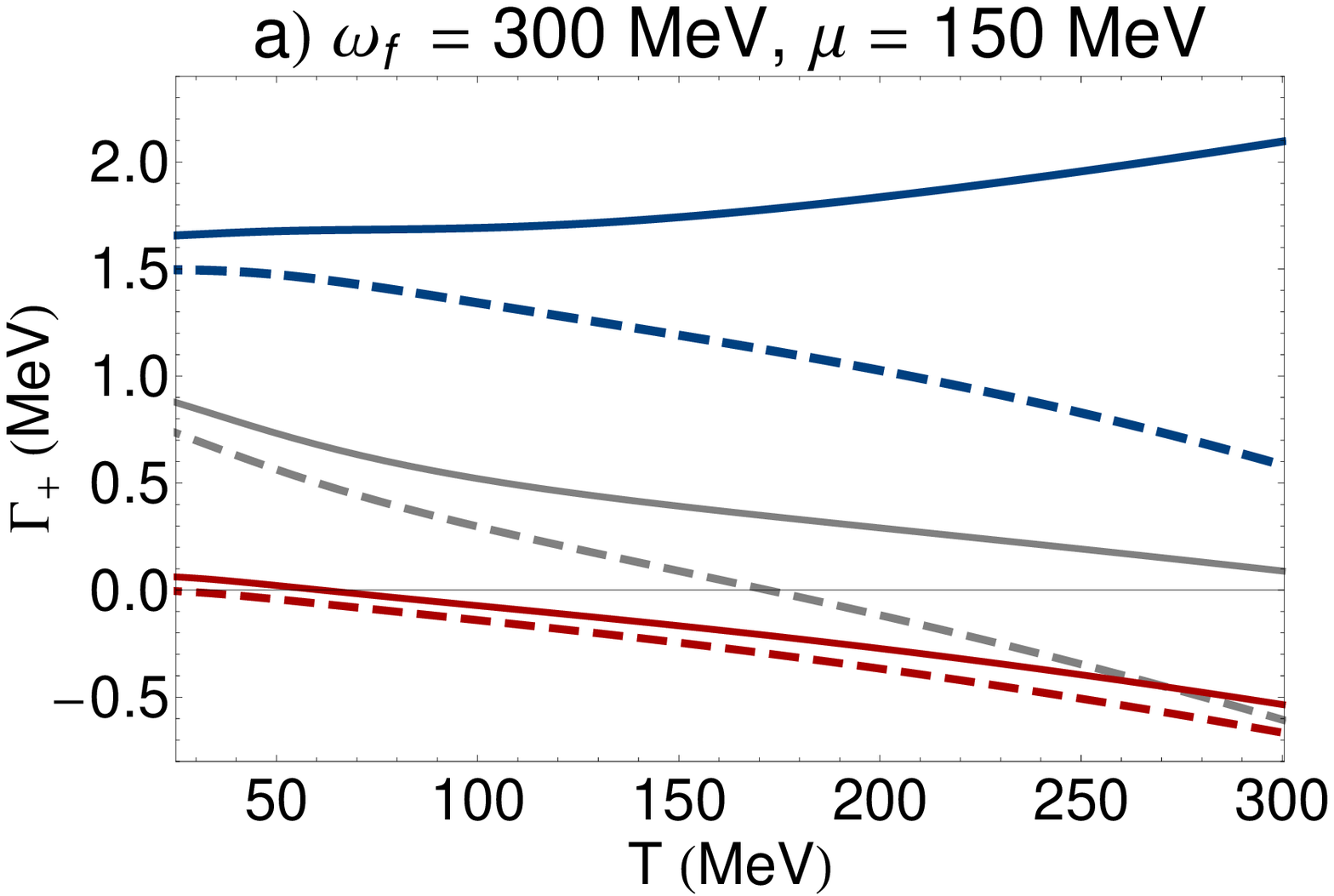}
\includegraphics[height=0.25\textheight]{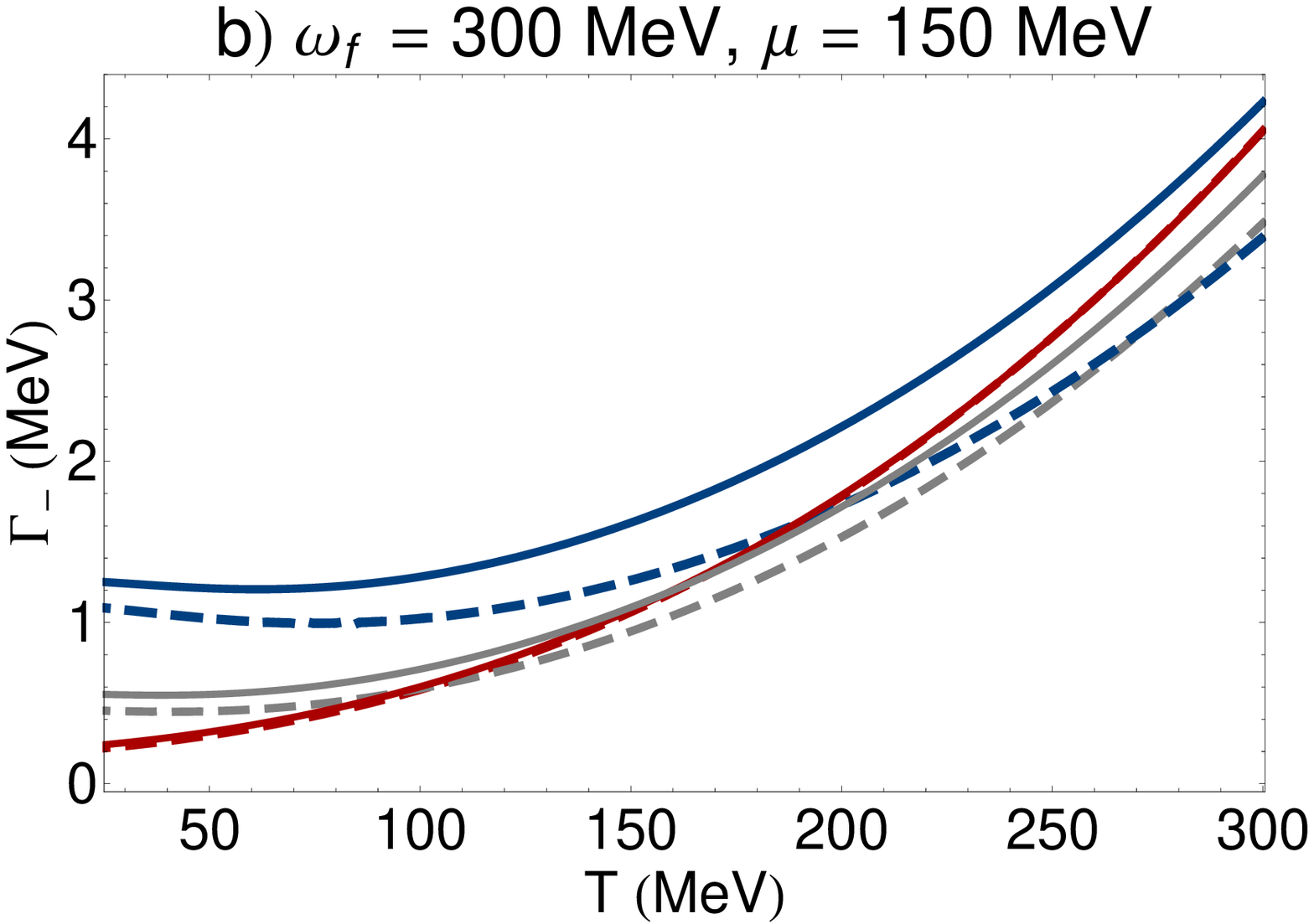}
\caption{(color online). The T dependence of (a) $\Gamma_{+}$, (b) $\Gamma_{-}$ is plotted for constant $\omega_{f}=300$ MeV and $\mu=150$ MeV. Dashed lines correspond to $\Gamma_{\pm}$, including (T, $\mu$)-independent $\xi_{0}=60,90,120$, while the solid lines correspond to the same quantities including the HTL corrections to the bosonic and fermionic masses~\ci{Taghinavaz2014}.}
\la{gammaft}
\end{figure}
\begin{figure}[!b]
\includegraphics[height=0.25\textheight]{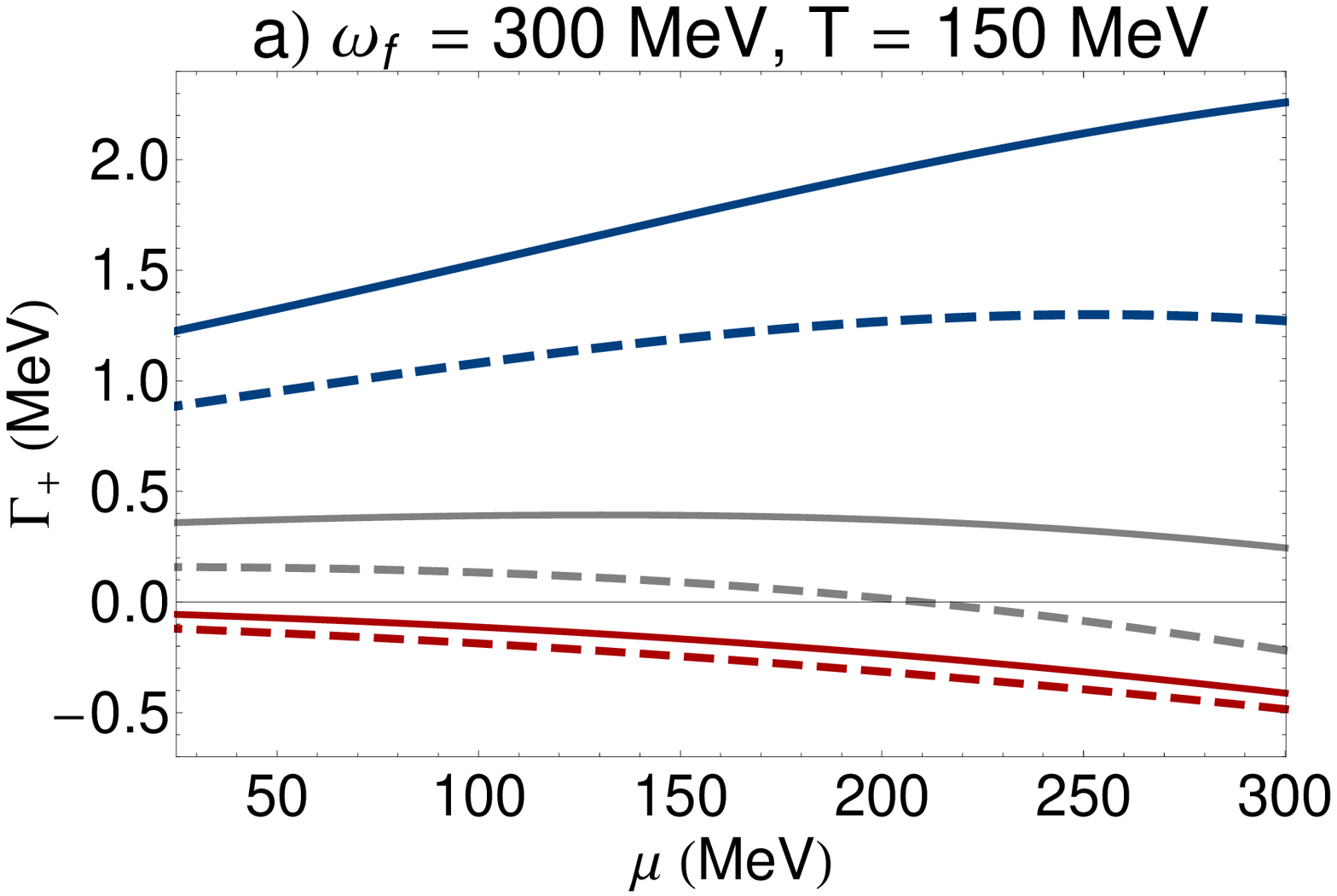}
\includegraphics[height=0.25\textheight]{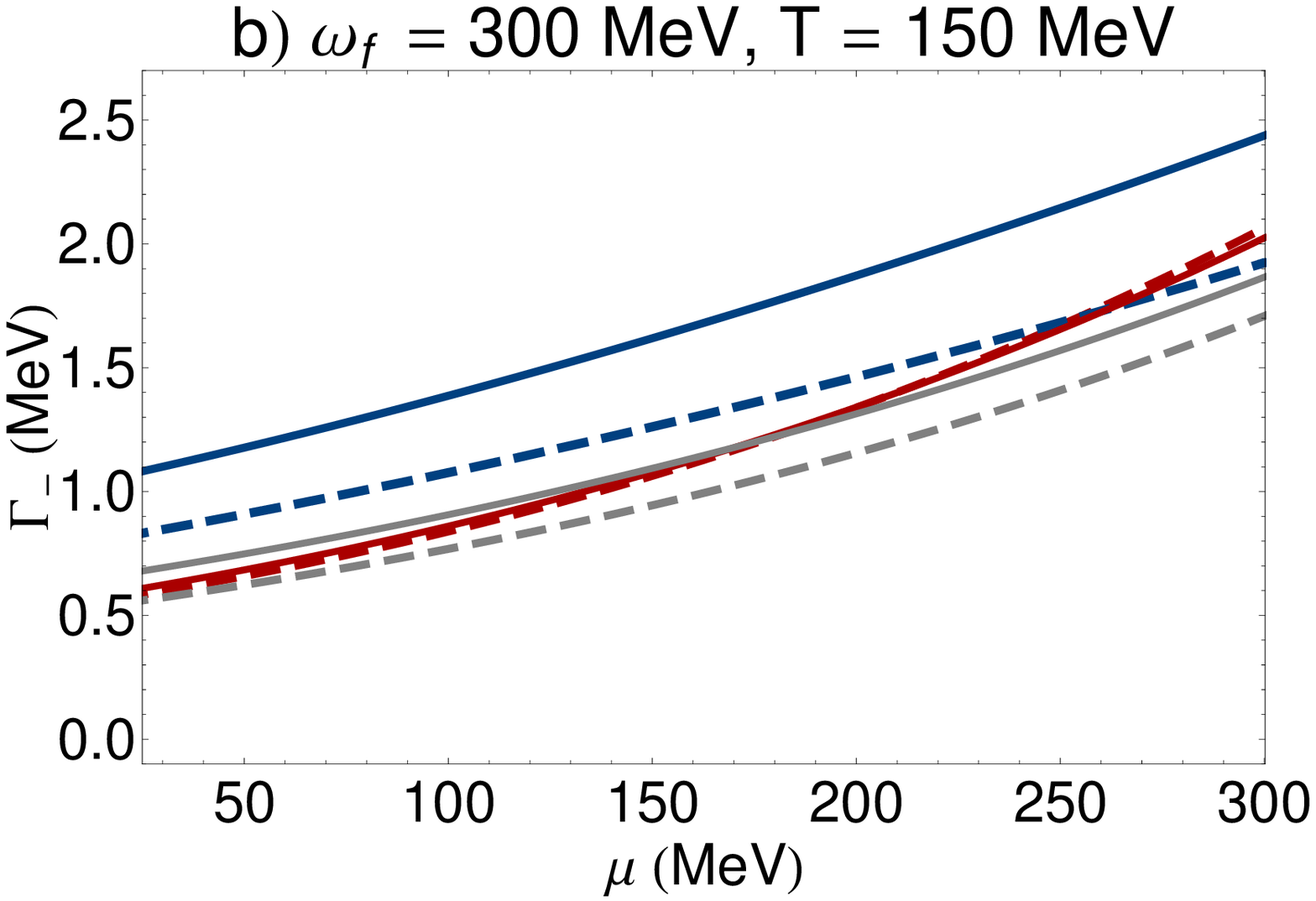}
\caption{(color online). The $\mu$ dependence of (a) $\Gamma_{+}$, (b) $\Gamma_{-}$ is plotted for constant $\omega_{f}=300$ MeV and $T=150$ MeV.  Dashed lines correspond to $\Gamma_{\pm}$, including (T, $\mu$)-independent $\xi_{0}=60,90,120$, while the solid lines correspond to the same quantities including the HTL corrections to the bosonic and fermionic masses~\ci{Taghinavaz2014}.}
\la{gammafmu}
\end{figure}
 \begin{figure}[!t]
\includegraphics[height=0.25\textheight]{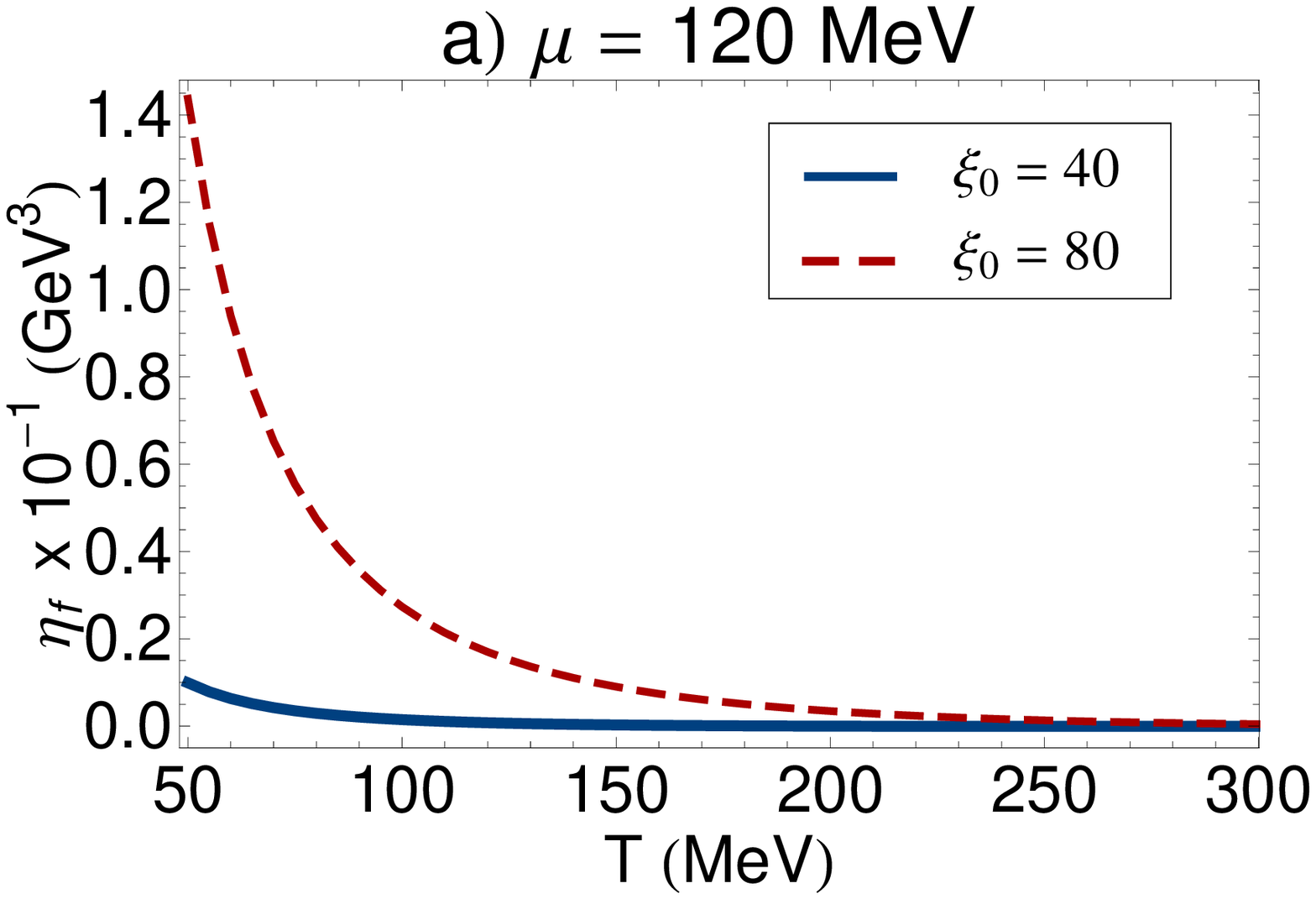}
\includegraphics[height=0.25\textheight]{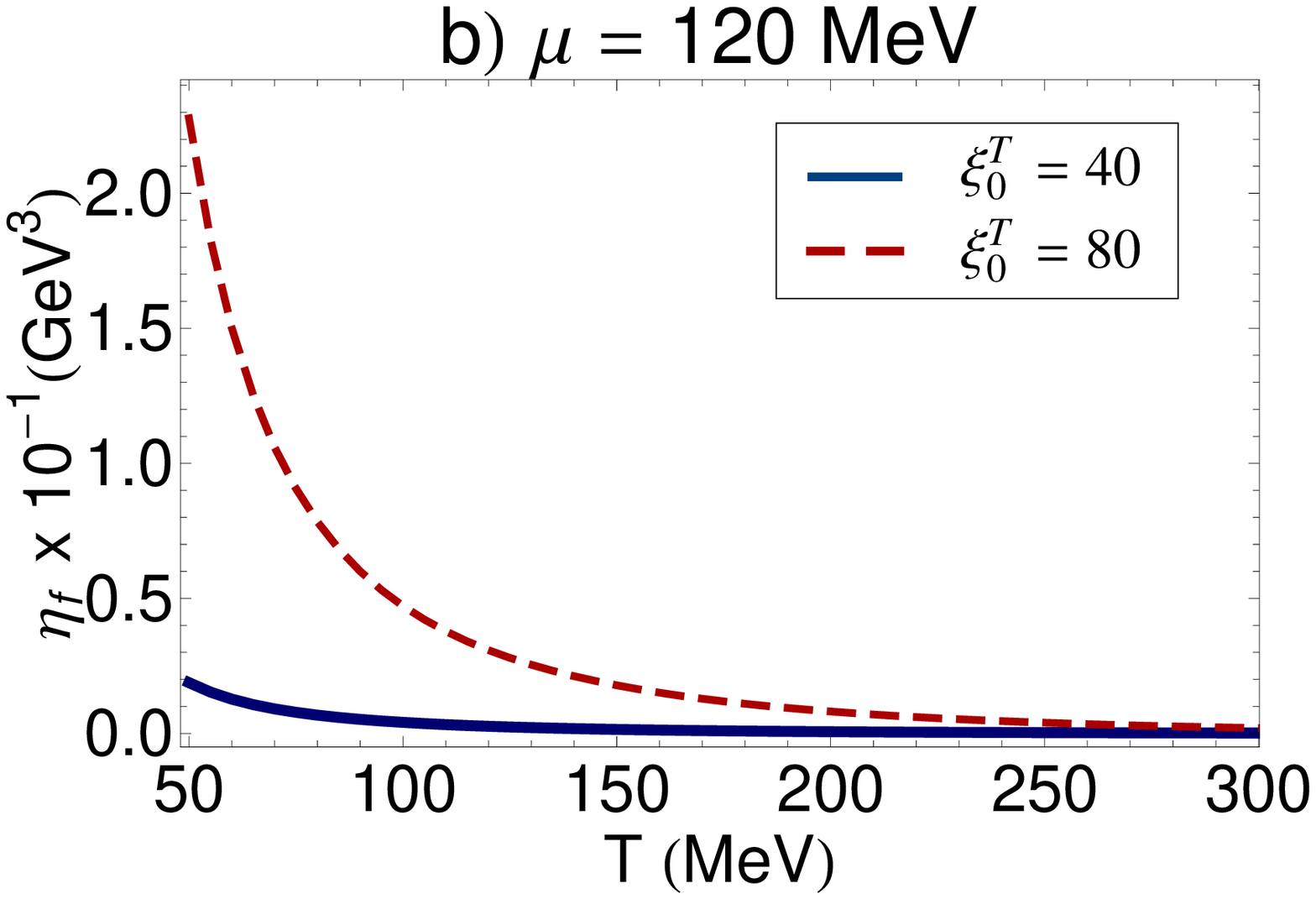}
\caption{(color online). The T dependence of $\eta_{f}$ is plotted for $\mu=120$ MeV and for (a) (T,$\mu$)-independent $\xi_{0}=40,80$ as well as (b) (T,$\mu$)-dependent $\xi_{0}^{T}=40,80$~\ci{Taghinavaz2014}.}
\la{etaft}
\end{figure}
\begin{figure}[!b]
\includegraphics[height=0.25\textheight]{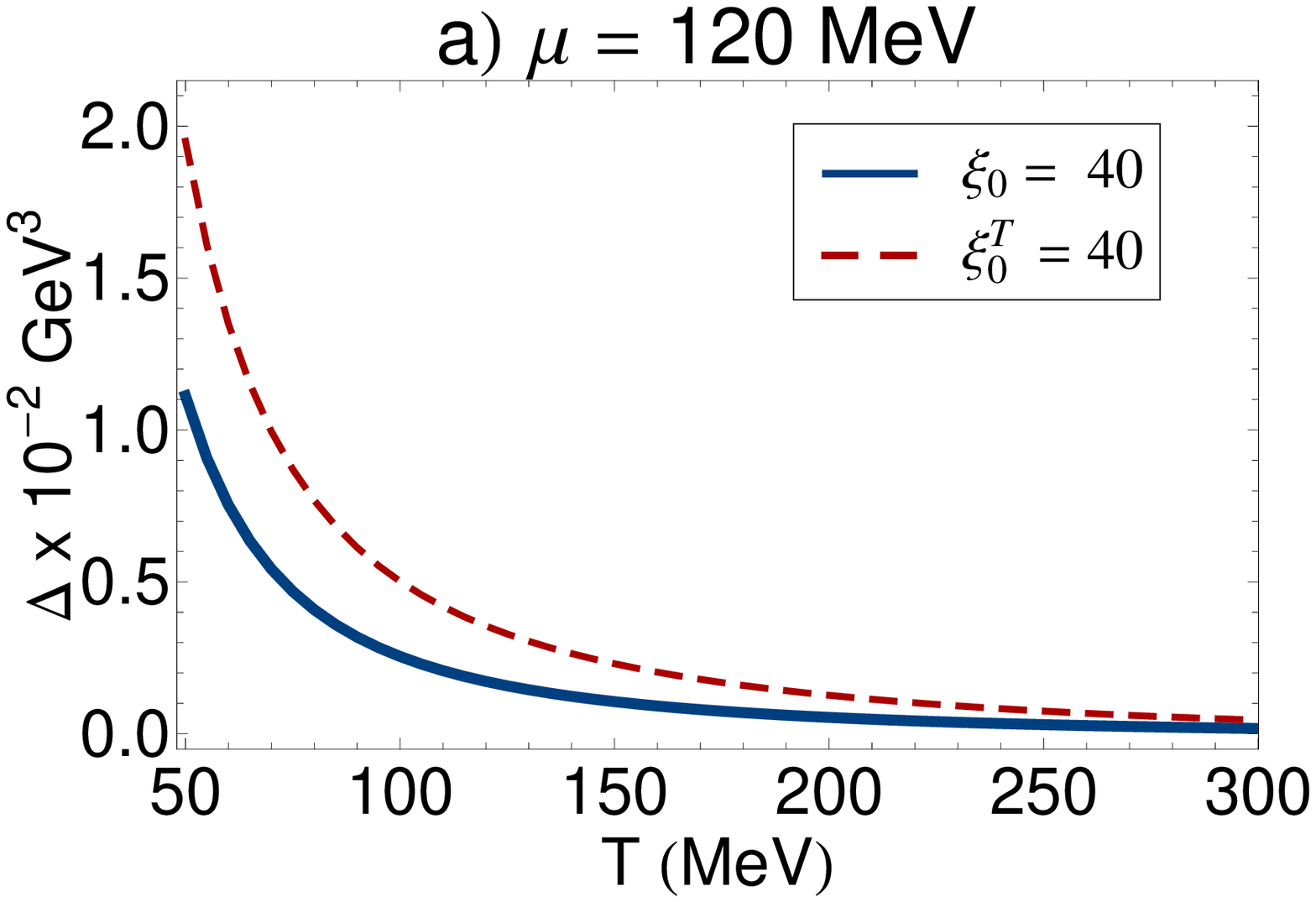}
\caption{(color online). The T dependence of $\Delta$ is plotted for $\mu=120$ MeV and $\xi_{0}=\xi_{0}^{T}=40$.}
\la{deltaeta}
\end{figure}
\se{Analytical Results}
According to the arguments presented in the previous sections, $\eta$ is sensitive to the low frequencies. By decomposing (\ref{21}) into bosonic and fermionic parts, we can calculate $\eta_{b}$ and $\eta_{f}$. The retarded bosonic and fermionic TPGF extracted from the first two terms of Fig. \ref{skeleton} read~\ci{Taghinavaz2014}
\beq
\Pi_{b}^{R}(p_0)&=& 4 \int \frac{d^{3}p}{(2\pi)^{3}} \eta^{\alpha \beta \rho \sigma} p_{\alpha}p_{\beta}p_{\rho} p_{\sigma} \int_{-\infty}^{\infty} \frac{d\omega_{1}~d\omega_{2}}{(2\pi)^{2}} \rho_{b} (\textbf{p},\omega_{1}) \rho_{b}(\textbf{p},\omega_{2})  n_{b}(\omega_{1}) n_{b}(\omega_{2}) W_{\epsilon}(\omega_{12},p_0),\\ \no
\Pi_{f}^{R}(p_0)&=& \frac{1}{(2\pi)^{2}}\int \frac{d^{3}p}{(2\pi)^{3}} \eta^{\alpha \beta \rho \sigma} p_{\rho} p_{\sigma}  \int_{-\infty}^{\infty} \frac{d\omega_{1}~d\omega_{2}}{(2\pi)^{2}} tr(\rho_{f} (\textbf{p},\omega_{1}) \gamma_{\alpha} \rho_{f}(\textbf{p},\omega_{2})\gamma_{\beta})  (1-n_{f}(\omega_{1})) n_{f}(\omega_{2}) W_{\epsilon}(\bar{\omega}_{12},p_0).\\ \no
\eeq
Here, $\eta^{\alpha \beta \rho \sigma}\equiv\Delta^{\nu \sigma}\Delta^{\mu \rho} + \Delta^{\nu \rho}\Delta^{\mu \sigma}- \frac{2}{3}\Delta^{\mu \nu}\Delta^{\rho \sigma}$, $\omega_{12}\equiv\omega_{1}+\omega_{2}$, $\bar{\omega}_{12}\equiv \omega_{1} - \omega_{2}$, and
$$W_{\epsilon}(\omega,p_0)\equiv\frac{1}{p_{0}+ i\epsilon - \omega} - \frac{1}{p_0 +i\epsilon + \omega}.$$
Moreover, $n_{b/f}(\omega)$  are bosonic ($b$) and fermionic ($f$) distribution functions, given by
\begin{eqnarray}
n_{b}(\omega)=\frac{1}{e^{\beta\omega}-1}, \qquad~n_{f}=\frac{1}{e^{\beta\omega}+1}.
\end{eqnarray}
The bosonic and fermionic spectral functions, $\rho_{b}(\textbf{p},\omega)$ and $\rho_{f}(\textbf{p},\omega)$, are defined in (\ref{27n}) and (\ref{24}), respectively. By inserting these relations into (\ref{23}), and after some straightforward computations, the bosonic and fermionic shear viscosity $\eta_{b}$ and $\eta_{f}$ read (see \cite{Taghinavaz2014} for more details)
\beq\la{311}
\eta_{b}&=& \frac{\beta}{30 \pi^{2}} \int_{0}^{\infty} dp \frac{\textbf{p}^{6}}{E_{b}^{2}} \frac{e^{\beta E_{b}}}{(e^{\beta E_{b}}-1)^{2}}\frac{1}{\Gamma_{b}} + \mathcal{O}(\Gamma_{b}),\nonumber\\
\eta_{f}&=& \frac{2\beta}{15 \pi^{2}} \int_{0}^{\infty} dp \frac{\textbf{p}^{4}}{\omega_{f}^{2}} \sum_{s=\pm} \bigg\{\frac{e^{\beta E_{s}}}{(e^{\beta E_{s}}+1)^{2}} \bigg[\frac{\textbf{p}^{2}}{\Gamma_{s}}- \frac{4m_{f}^{2} (\Gamma_{f}^{+}- \Gamma_{s})}{[E_{f}+ is \Gamma_{f}^{+}] [E_{f}+ i\Gamma_{f}^{-}]}\bigg]\bigg\} + \mathcal{O}(\Gamma_{\pm}).
\eeq
In the fermionic channel, $s=\pm$ refer to the particle ($s=+$) and plasmino ($s=-$) excitations, respectively. In addition, $\Gamma_{f}^{\pm}$ are defined by
$\Gamma_{f}^{\pm}\equiv\Gamma_{+} \pm \Gamma_{-}$.
We are working in a frame, in which quantum fluctuations do not imply any sensible effects on dispersion relations. We therefore adopt the approximations $E_{b}\approx\omega_{b}$ and $E_{f}\approx\omega_{f}$, and compute $\Gamma_{b}$ as well as $\Gamma_{\pm}$ using the RTF of FTFT up to first order correction of a perturbative expansion in the orders of the Yukawa coupling constant $g$. Analytical results for the bosonic and fermionic decay widths are given by
\beq
\Gamma_{b}(\textbf{p},\omega)=\frac{g^{2}T}{16 \pi} \frac{\gamma_{b}^{2} (\xi^{2}-4)}{\xi^{2} \sqrt{1- \gamma_{b}^{2}}} \ln\bigg[ \frac{\cosh(\tau_{f}) + \cosh\frac{\kappa_{b}}{2}\bigg(1+ \frac{1}{\xi}\sqrt{(\xi^{2}- 4) (1-\gamma_{b}^{2})}\bigg)}{\cosh(\tau_{f}) + \cosh\frac{\kappa_{b}}{2}\bigg(1 - \frac{1}{\xi}\sqrt{(\xi^{2}- 4) (1-\gamma_{b}^{2})}\bigg)}\bigg],
\eeq
and
\beq
\Gamma_{+}(\textbf{p},\omega)&=&\frac{g^{2}T}{32 \pi} \frac{\gamma_{f}^{2} (\xi^{2}-4)}{\sqrt{1- \gamma_{f}^{2}}} \bigg\{\ln\bigg[ \frac{1 - \cosh(2\Xi_{-})}{\cosh(\Upsilon_{-}+\Xi_{+}) - \cosh(\Upsilon_{-} -\Xi_{+})}\bigg] \nonumber\\
&&- \ln\bigg[ \frac{1 +\cosh(2\Xi_{-}- (\kappa_{f}+ \tau_{f}))}{\cosh(\Upsilon_{-}+\Xi_{+}) - \cosh(\Upsilon_{+} -\Xi_{+} + \tau_{f})}\bigg]\bigg\}.
\eeq
Moreover, we have $\Gamma_{-}=\Gamma_{+} - \Gamma_{f}^{-}$ with
\beq
\Gamma_{f}^{-}&=&- \frac{g^{2}T}{8 \pi \kappa_{f}\sqrt{1 - \gamma_{f}^{2}}} \bigg\{ \kappa_{f} \ln\bigg[ \frac{1 - \cosh(2\Xi_{-})}{\cosh(\Upsilon_{-}+\Xi_{+}) - \cosh(\Upsilon_{-} -\Xi_{+})}+ \tau_{f} \ln\bigg[ \frac{1 +\cosh(2\Xi_{-}- (\kappa_{f}+ \tau_{f}))}{\cosh(\Upsilon_{-}+\Xi_{+}) - \cosh(\Upsilon_{+} -\Xi_{+} + \tau_{f})}\nonumber\\
&& + [u(u + 2\ln(1 - e^{-2u})) - \mbox{Li}_{2}(e^{-2u})]|_{\Upsilon_{-}}^{- \Xi_{-}} + [u(u + 2\ln(1 - e^{-2u})) - \mbox{Li}_{2}(e^{-2u})]|_{\Xi_{-}}^{ \Xi_{+}}\nonumber\\
&& -  [u(u + 2\ln(1 + e^{-2u})) - \mbox{Li}_{2}(-e^{-2u})]|_{\Upsilon_{-}+ \frac{\kappa_{f}+\tau_{f}}{2}}^{- \Xi_{-} + \frac{\kappa_{f}+\tau_{f}}{2}} - [u(u + 2\ln(1 + e^{-2u})) - \mbox{Li}_{2}( - e^{-2u})]|_{\Xi_{-} -  \frac{\kappa_{f}+\tau_{f}}{2}}^{\Xi_{+} - \frac{\kappa_{f}+\tau_{f}}{2}} \bigg\}.
\eeq
In the above relations, $\gamma_{b}\equiv m_{b}/\omega_{b},  \gamma_{f}\equiv m_{f}/\omega_{f}$ and $\xi\equiv m_{b}/m_{f}$ as well as $\tau_{f}\equiv \mu\beta, \kappa_{f}\equiv \omega_{b}\beta$ and $\kappa_{f}\equiv \omega_{f}\beta$.
We have also used the definitions
$$\Xi_{\pm}\equiv\kappa_{f} \xi \big[\xi \pm \sqrt{(\xi^{2} - 4) (1- \gamma_{f}^{2})}\big]/2,\qquad \mbox{and }\qquad\Upsilon_{\pm} \equiv\kappa_{f}(\gamma_{f}\pm 1)/2.$$
To consider HTL corrections for bosonic and fermionic masses, we have introduced $\xi^{T}=m_{b}(T,\mu)/m_{f}(T,\mu)$ with $m_{b}(T,\mu)\equiv m_{b}^{0} + m_{b}^{th}(T,\mu)$ and $m_{f}(T,\mu)\equiv m_{f}^{0} + m_{f}^{th}(T,\mu)$. Here, the superscript $0$ corresponds to the T- and $\mu$-independent bosonic and fermionic masses, whose thermal corrections are given by
\beq \la{31}
m_{b}^{th}(T,\mu)&=& \frac{g^{2}}{6} (T^{2} + \frac{3\mu^{2}}{\pi^{2}}),\nonumber\\
m_{f}^{th}(T,\mu)&=& \frac{g^{2}}{16} (T^{2} + \frac{\mu^{2}}{\pi^{2}}).
\eeq
Let us notice at this stage that the coupling $g$ does not receive any HTL corrections in the Yukawa theory.  In this section, we have presented the basic analytical expressions, which are necessary for calculating the decay widths $\Gamma_{b}$ and $\Gamma_{\pm}$ as well as the shear viscosities $\eta_{b}, \eta_{f}$ in terms of free parameters of the theory. The T and $\mu$ dependence of the above quantities will be demonstrated in the next section.
\se{Numerical Results}
The T and $\mu$ dependence of $\Gamma_{b}$ is demonstrated in Fig. \ref{gammabt}.
As it turns out, by increasing T and $\mu$, $\Gamma_{b}$ decreases. This reflects the fact that the bosonic particles have a larger mean free path in dense or hot medium.  By considering thermal corrections to the bosonic mass, the mean free path of bosons increases. The T and $\mu$ dependence of $\eta_{b}$ can be determined by inserting the expression for  $\Gamma_{b}$ into the (\ref{311}).  In Fig. \ref{etabt}, the T dependence of $\eta_{b}$ is demonstrated for $\mu=120$ MeV and for T- and $\mu$-independent and dependent bosonic to fermionic mass ratios, $\xi_{0}$ and $\xi_{0}^{T}$. The fact that $\eta_{b}$ increases with $T$ demonstrates the fact that the mean free path (time) of bosons increases in a hot medium. This is consistent with the arguments based on relaxation time approximation. By including HTL corrections to the particle masses at fixed $\mu$, $\eta_{b}$ decreases for each fixed T and $\mu$. 
\par
In Figs. \ref{gammaft} and \ref{gammafmu}, the T and $\mu$ dependence of the decay widths corresponding to particle and plasmino modes are plotted.
The T and $\mu$ dependence of $\Gamma_{+}$ is somehow similar to that corresponding to $\Gamma_{b}$. In contrast, the T and $\mu$ dependence of $\Gamma_{-}$ seems to be different, in the sense that by increasing T and $\mu$, $\Gamma_{-}$ increases. This indicates that at high temperature or in dense medium the plasmino modes wash out earlier than the particle modes. In particular, this means that by increasing T and $\mu$, plasmino modes decay into the normal particle modes, and hence the normal particle lifetime increases with increasing temperature or chemical potential.
By plugging $\Gamma_{\pm}^{f}$ into (\ref{311}), the T dependence of $\eta_{f}$ can be determined for different $\xi_{0}=40,80$ and $\xi_{0}^{T}=40,80$ and fixed $\mu$. The results of $\eta_{f}$ are demonstrated in Fig. \ref{etaft}. Unlike the bosonic case,  the fermionic shear viscosity is a decreasing function in terms of T. This denotes that the mean free path (time) of fermionic particles decrease with increasing temperature. Moreover, by including HTL corrections for particle masses, $\eta_{f}$ increases for each fixed T and $\mu$, in contrast to the bosonic case.
\par
Our main result is presented in Fig. \ref{deltaeta}, where the T dependence of a certain quantity $\Delta$ is presented. Here, $\Delta$ is defined by the difference between $\eta_{f}$ in terms of $\Gamma_{+}=\Gamma_{-}$ and $\Gamma_{+}\neq\Gamma_{-}$, i.e.,
\beq
\Delta\equiv\eta_{f}[\Gamma_{+}=\Gamma_{-}] - \eta_{f}[\Gamma_{+}\neq \Gamma_{-}].
\eeq
Figure \ref{deltaeta} illustrates the fact that by increasing temperature, the effect of plasmino modes on $\eta_{f}$ become negligible, and the approximation $\Gamma_{+}\approx \Gamma_{-}$ is therefore reliable only in the limit of high temperature~\ci{Gagnon2007, Basagoiti2002}.
\se{Concluding remarks}
The shear viscosity is a transport coefficient which quantifies the long wave-length response of a medium to momentum anisotropies. In the present work, we 
calculated the bosonic and fermionic parts of the shear viscosity of a hot and dense Yukawa-Fermi gas in the leading ${\cal{O}}(\Gamma^{-1})$ order. 
Since our theory is massive, the small-angle scattering divergences are irrelevant, and a perturbative computation of transport coefficients is reliable. 
We showed that by increasing the temperature, $\eta_{b}$ increases. In contrast, $\eta_{f}$ decreases with increasing temperature. 
We also took the HTL corrections to particle masses into account and investigated their effects on $\eta_{b}$ as well as $\eta_{f}$. We eventually showed 
that the assumption $\Gamma_{+}\approx\Gamma_{-}$ is only justified at high temperature. 
Here, $\Gamma_{+}$ and $\Gamma_{-}$ are the decay widths corresponding to particle and plasmino excitations.



\bibliographystyle{aipproc}   

\bibliography{sample}

\IfFileExists{\jobname.bbl}{}
 {\typeout{}
  \typeout{******************************************}
  \typeout{** Please run "bibtex \jobname" to optain}
  \typeout{** the bibliography and then re-run LaTeX}
  \typeout{** twice to fix the references!}
  \typeout{******************************************}
  \typeout{}
 }


\end{document}